\documentclass[sigconf]{acmart}
\usepackage{multirow}
\AtBeginDocument{%
  \providecommand\BibTeX{{%
    \normalfont B\kern-0.5em{\scshape i\kern-0.25em b}\kern-0.8em\TeX}}}

\copyrightyear{2021}
\acmYear{2021}
\setcopyright{acmlicensed}
\acmConference[CIKM '21]{Proceedings of the 30th ACM International Conference on Information and Knowledge Management}{November 1--5, 2021}{Virtual Event, QLD, Australia}
\acmBooktitle{Proceedings of the 30th ACM International Conference on Information and Knowledge Management (CIKM '21), November 1--5, 2021, Virtual Event, QLD, Australia}
\acmPrice{15.00}
\acmDOI{10.1145/3459637.3482489}
\acmISBN{978-1-4503-8446-9/21/11}
\begin{document}
\fancyhead{}
\title{USER: A Unified Information Search and Recommendation Model based on Integrated Behavior Sequence}

\author{Jing Yao$^{1,2}$, Zhicheng Dou$^{1*}$, Ruobing Xie$^{2}$, Yanxiong Lu$^{2}$, Zhiping Wang$^{2}$ and Ji-Rong Wen$^{3,4}$}
\affiliation{$^1$Gaoling School of Artificial Intelligence, Renmin University of China}
%\affiliation{$^2$School of Information, Renmin University of China}
\affiliation{$^2$WeChat Search Application Department, Tencent, China}
\affiliation{$^3$Beijing Key Laboratory of Big Data Management and Analysis Methods}
\affiliation{$^4$Key Laboratory of Data Engineering and Knowledge Engineering, MOE}
\email{{jing_yao,dou*}@ruc.edu.cn}

\begin{abstract}
Search and recommendation are the two most common approaches used by people to obtain information. They share the same goal -- satisfying the user's information need at the right time. There are already a lot of Internet platforms and Apps providing both search and recommendation services, showing us the demand and opportunity to simultaneously handle both tasks. However, most platforms consider these two tasks independently -- they tend to train separate search model and recommendation model, without exploiting the relatedness and dependency between them.
In this paper, we argue that jointly modeling these two tasks will benefit both of them and finally improve overall user satisfaction. We investigate the interactions between these two tasks in the specific information content service domain. We propose first integrating the user's behaviors in search and recommendation into a heterogeneous behavior sequence, then utilizing a joint model for handling both tasks based on the unified sequence. More specifically, we design the \textbf{U}nified Information \textbf{SE}arch and \textbf{R}ecommendation model (\textbf{USER}), which mines user interests from the integrated sequence and accomplish the two tasks in a unified way.
%Our unified model has several advantages: (1) Aggregating search and recommendation logs together can alleviate the problem of data sparsity. (2) More accurate user interests can be learned from integrated behavior sequences. (3) We can capture the potential relatedness between the two tasks to essentially promote each other.
Experiments on a dataset from a real-world information content service platform verify that our model outperforms separate search and recommendation baselines.
\end{abstract}

\begin{CCSXML}
<ccs2012>
<concept>
<concept_id>10002951.10003260.10003261.10003263</concept_id>
<concept_desc>Information systems~Web search engines</concept_desc>
<concept_significance>500</concept_significance>
</concept>
<concept>
<concept_id>10002951.10003317.10003331.10003271</concept_id>
<concept_desc>Information systems~Personalization</concept_desc>
<concept_significance>500</concept_significance>
</concept>
<concept>
<concept_id>10002951.10003317.10003347.10003350</concept_id>
<concept_desc>Information systems~Recommender systems</concept_desc>
<concept_significance>500</concept_significance>
</concept>
</ccs2012>
\end{CCSXML}

\ccsdesc[500]{Information systems~Web search engines}
\ccsdesc[500]{Information systems~Personalization}
\ccsdesc[500]{Information systems~Recommender systems}

\keywords{personalized search; recommendation; unified model}
\def\authors{Jing Yao, Zhicheng Dou, Ruobing Xie, Yanxiong Lu, Zhiping Wang, and Ji-Rong Wen}
\maketitle

\begin{figure}
    \centering
    \includegraphics[width=0.99\linewidth,height=5.05cm]{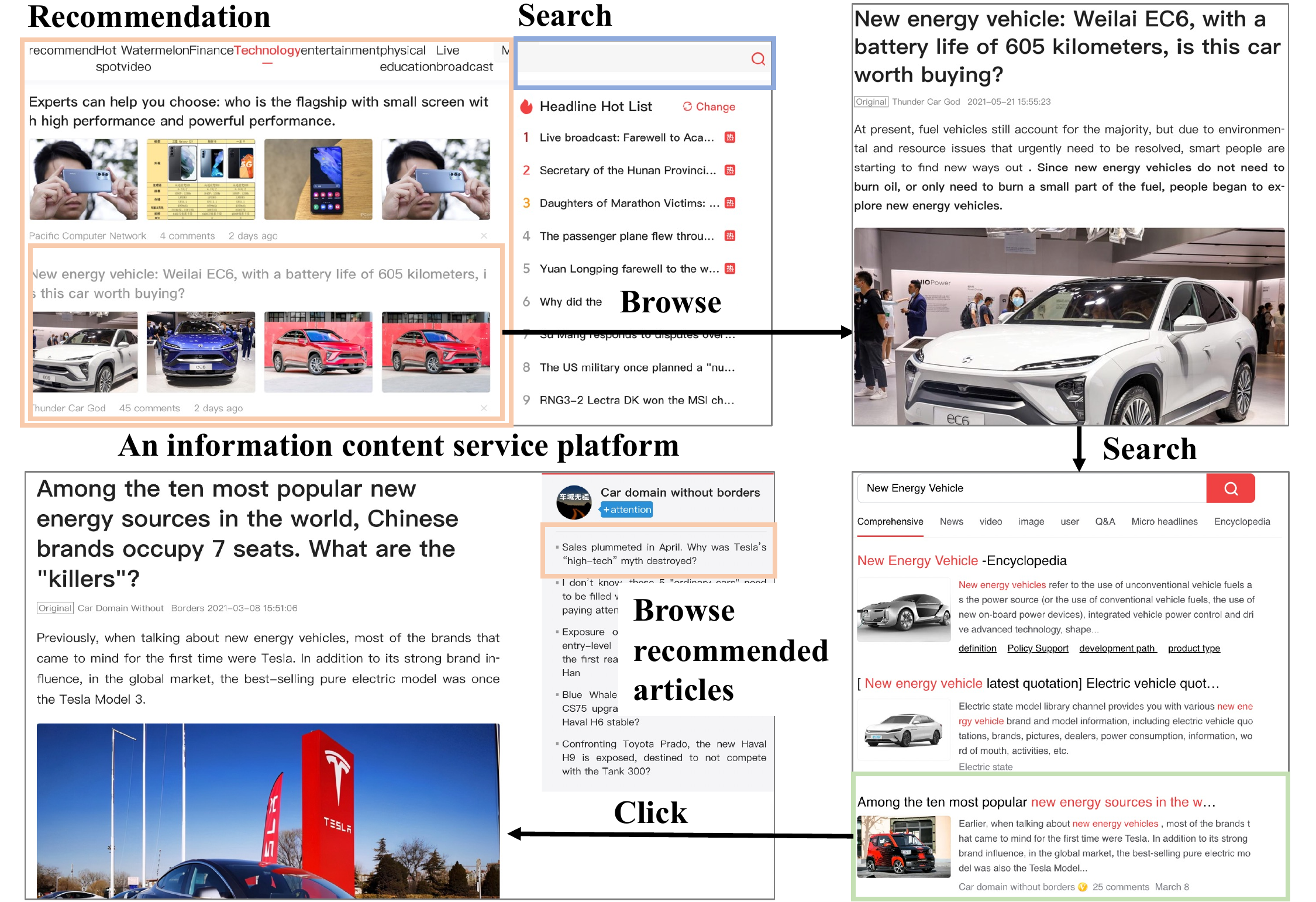}
    \caption{Illustration of an information content service platform with both search and recommendation services.}
    \label{fig:example}
\end{figure}

\section{Introduction}\label{sec:intro}
On Internet platforms, search and recommendation are two major approaches to help users obtain the required knowledge. In this paper, we mainly focus on the domain of information content service which aims to deliver news feeds, tweets, or web articles to users\footnote{Example information content service platforms are twitter, toutiao, and wechat.}. In order to improve users' satisfaction with search and recommendation results, a lot of personalized search models and recommendation models have been proposed~\cite{Bennett_2012_SLTB,Ge_2018_HRNN,Lu_2019_PSGAN,Cheng_2016_WideDeep,Wu_2019_NPA,Wu_2019_NAML,Ge_2020_GERL}. These models aim to mine user preferences from their historical behaviors to infer their current intents and generate a personalized document ranking list that can satisfy the current user interest. Typically, many deep learning based personalized search models learn a representation of user interests from her search history to re-rank the candidate documents\cite{Ge_2018_HRNN,Lu_2019_PSGAN,Yao_2020_RLPer,Lu_2020_KEPS,Yao_2020_PEPS,Zhou_2020_HTPS}. Recommendation models also present document ranking lists according to the user's browsing history~\cite{Wu_2019_NPA,Wu_2019_NAML,Wu_2019_NRMS,Ge_2020_GERL,Xie_2020_kanyikan}. However, most existing studies concentrate on only one single task, namely either search or recommendation. They devise a specific model applicable for one task, but rarely consider their combination.

Currently, there are more and more mobile Apps and websites where both information search and recommendation services are available. For the example of Toutiao\footnote{https://www.toutiao.com/} platform shown in Figure~\ref{fig:example}, users can not only actively issue queries to seek information, but also browse the recommended articles. Indeed, some early attempts of combining the two services have already been applied. For example, some articles are recommended along with the clicked search results. Queries may also be suggested at the end of a recommended news article. \textbf{Therefore, how to effectively aggregate the two tasks together is an essential and valuable problem.}

Actually, some early studies~\cite{Belkin_1992_earlydiscussion} have discussed the similarity between search and recommendation. The two tasks share the same target -- helping people get the information they require at the right time. \citet{Zamani_2018_JSR} propose a vanilla joint learning framework to handle both tasks at the same time. They train two separate models for the two tasks through a joint loss, but neglect the essential relatedness between them in human information-seeking behaviors. Actually, users usually switch between the two services when they are obtaining information from the Web. Let us take the example in Figure~\ref{fig:example} for illustration. When a user browses the article list generated by the recommendation system, she is attracted by the article titled ``New energy vehicle: Weilai ...?''. After reading this article, she switches to the search engine and issues a query to seek more knowledge about ``New energy vehicle''. Then, she browses the search results and articles recommended along with the clicked document to know more. Such an information-seeking pattern which mixes behaviors made in proactive searches and passive recommendations is common in our surfing process. From the example, we find that the user may switch between the search service and recommender system for a single target, both the search behaviors and browsing behaviors reflect her personalized information need. Therefore, jointly modeling the entire user behavior sequence is expected to discover real user intents more precisely. Besides, some close associations may exist between the two kinds of behaviors that browsing could stimulate search and search might impact browsing in the future. Richer interaction and training data is available. Motivated by this scenario, \textbf{we pay attention to jointly modeling both tasks of personalized search and recommendation in the information content domain, exploring the potential relatedness between their corresponding user behaviors to promote each other.}

To begin with, we integrate the user's historical search and browsing behaviors in chronological order, getting a simplified heterogeneous behavior sequence shown in Figure~\ref{fig:behavior_sequence}. $B$ represents browsed articles, $Q$ indicates queries issued by the user, and $D$ is documents clicked under the corresponding query. Then, we propose a \textbf{U}nified Information \textbf{SE}arch and \textbf{R}ecommendation model (\textbf{USER}) to encode the heterogeneous sequence and solve the two tasks in a unified way. We think recommendation and personalized search share the same paradigm: recommendation can be treated as personalized search taking an \texttt{EMPTY} query. Hence we design the USER model in a personalized ranking style, to rank candidate documents based on the input query (using empty for recommendation) and the user preferences contained in the integrated behavior sequence. This model has several advantages. \textbf{First}, we aggregate the user's search and recommendation logs, alleviating the problem of data sparsity faced by a single task. \textbf{Second}, based on the merged behavior sequence, more comprehensive and accurate user profiles can be constructed, improving personalization performance. \textbf{Third}, the potential relatedness between search and recommendation can be captured to essentially promote each other.

Specifically, our \textbf{USER} model is composed of four modules. \textbf{First}, a text encoder is used to learn the representation for the documents and queries. \textbf{Second}, the session encoder models the integrated behavior sequence in the current session, captures relatedness between the search and browsing behaviors, and clarifies the user's current intention. As for a search behavior including a query and clicked documents with strong relevance, we employ a co-attention structure~\cite{Shu_2019_CoAttention} to fuse their representations. Then, a transformer layer is constructed to capture the associations between the search and browsing behaviors in the session and fuse the context into the current intention. \textbf{Third}, the history encoder learns information from the long-term heterogeneous history sequence as an enhancement. \textbf{Finally}, we build a unified task framework to complete the two tasks in a unified way. We first pre-train the unified model with the training data from both tasks, alleviating data sparsity. Then, we make a copy for each task and finetune it with the corresponding task data to fit the individual data distribution. We experiment on a dataset comprised of search and browsing behaviors constructed from a real-world information content service platform with both search and recommendation engines. The results verify that our model outperforms separate baselines and alleviates data sparsity.

Our main contributions are summarized as follows: (1) We pay attention to both tasks of personalized search and recommendation. For the first time, we integrate separate behaviors of the two tasks into a heterogeneous behavior sequence. (2) We model the relatedness between a user's search and browsing behaviors to promote both personalized search and recommendation. (3) We propose a unified search and recommendation model (USER) that accomplishes the two tasks in a unified way with an encoder for the integrated behavior sequence and a unified task framework. 

\section{Related Work}\label{sec:related_work}
\subsubsection*{Personalized Search}
Personalized search customizes search results for each user by inferring her personal intents. Early studies relied on features and heuristic methods to analyze user interests. Focusing on click features, Dou et al.~\cite{Dou_2007_PClick} proposed P-Click to re-rank documents with their historical click counts. Topic-based features were applied to build user profiles~\cite{Sieg_2007_ODP,Bennett_2010_ODP,White_2013_ODP,Carman_2010_LDA,Harvey_2013_LDA,Vu_2015_LDA,Vu_2017_LDA}. The Open Directory Project (ODP)~\cite{Sieg_2007_ODP}, learned or latent topic models~\cite{Blei_2001_LDA} were used to obtain the topic-based information of a web page. Besides, the user's reading level and location are applied for personalization~\cite{Bennett_2011_readinglevel,Bennett_2011_location}. Multiple features were combined with a learning to rank method~\cite{Burges_2005_lambdarank,burges_2008_lambdaMart} to compute a personalized score~\cite{Bennett_2012_SLTB,Volkovs_2015_features}.

Recently, deep learning was applied to capture potential user preferences. Song et al.~\cite{2014_Song_AdaptingRanknet} leveraged personal data to adapt a general ranking model. Ge et al.~\cite{Ge_2018_HRNN} devised a hierarchical RNN with query-aware attention to dynamically mine preference information. Lu et at.~\cite{Lu_2019_PSGAN} employed GAN~\cite{Goodfellow_2014_GAN} to enhance the training data. Yao et al.~\cite{Yao_2020_RLPer} adopted reinforcement learning to learn user interests. Zhou et al.~\cite{Zhou_2020_RPMN} explored re-finding behaviors with a memory network. The latest studies were committed to disambiguating the query by introducing entities~\cite{Lu_2020_KEPS}, training personal word embeddings~\cite{Yao_2020_PEPS}, or involving search history as the context~\cite{Zhou_2020_HTPS}. All these models are specially designed for the personalized search task.

\subsubsection*{Information Recommendation Models}
Personalized content recommendation is critical to help users alleviate information overload and find something interesting. Traditional recommendation systems mainly depended on collaborative filtering (CF)~\cite{Sarwar_2001_CF} and factorization machine (FM)~\cite{Rendle_2010_FM}. With the emergence of deep learning, many models combined both low- and high-order feature interactions, such as Wide \& Deep~\cite{Cheng_2016_WideDeep} and DeepFM~\cite{Guo_2017_DeepFM}. Specially, representation based models have been studied for the recommendation of news articles that have abundant textual information. These models include two modules: a text encoder to obtain article representations and a user encoder to learn user representation from her browsing history. Then, articles are ranked based on their relevance with the user. Okura et al~\cite{Okura_2017_autoencoder} devised an auto-encoder to learn news representations, and used an RNN to generate user representations. Wu et al.~\cite{Wu_2019_NAML} learned article vectors from titles, bodies and topic categories. User representation was a weighted sum of the browsed news vectors. Wu et al.~\cite{Wu_2019_NPA} set user embeddings to generate personalized attention to calculate the article and user representations. They also exploited multi-head self-attention~\cite{Vaswani_2017_Tranformer} to capture contextual information~\cite{Wu_2019_NRMS}. LSTUR~\cite{An_2019_LSTUR} kept both short-term and long-term user profiles. To enhance text representations, entities in the article and their neighbors in the knowledge graph are considered~\cite{Wang_2018_DKN,Wang_2019_Knowledge,Liu_2020_knowledge}. The GNN structure~\cite{Wu_2019_GNNSurvey} was also adopted to model high-order relatedness between users and articles~\cite{Hu_2020_Graph,Ge_2020_GERL}. In these models, only the recommendation task is discussed.

\subsubsection*{Joint Search and Recommendation}
Some studies considered both the search and recommendation tasks. In e-commerce, an early work~\cite{Wang_2012_USREcommerce} built a unified recommendation and search system by merging their features. Zamani et al.~\cite{Zamani_2018_JSR} proposed a joint learning framework that simultaneously trains a search model and a recommendation model by optimizing a joint loss. For the situation with only recommendation data but not search logs, a multi-task framework was trained on browsing interactions~\cite{Zamani_2020_JSR}. These joint methods simply combined the two tasks and train two separate models through multi-task learning or joint loss, without exploring more essential dependency between them. Search history was also used to help generate recommendations for the users with little browsing history~\cite{Wu_2019_HeterogeneousBehavior,Yao_2012_SearchHelpReco}. This model just targeted one single task with data from the other task as complementary information. In this paper, we propose a unified model to solve the two tasks at the same time, mining the relatedness between their corresponding user behaviors to promote each other.
\begin{figure}
    \centering
    \includegraphics[width=0.95\linewidth]{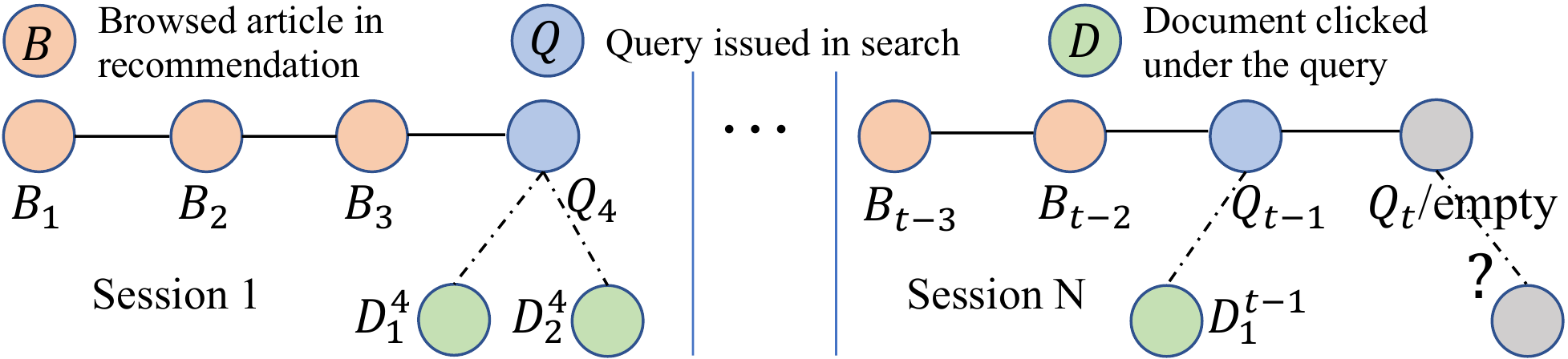}
    \caption{Illustration of the integrated behavior sequence. The target behavior is personalized search with a query $Q_t$ or recommendation with an empty query.}
    \label{fig:behavior_sequence}
\end{figure}

\section{Problem Definition}\label{sec:definition}
Search and recommendation are two main approaches to help people obtain information. Many separate personalized search models and recommendation models have been proposed. As analyzed in Section~\ref{sec:intro}, people usually achieve their information targets through a mixture of proactive searches and passive recommendation, which is popular on information content service platforms with both search and recommendation engines. Both kinds of behaviors reflect the user's information need and preferences. Thus, compared to existing separate approaches, jointly modeling the two tasks and exploiting the relatedness between them might have the potential to promote each other. In this paper, we integrate the user's search and browsing behaviors into a sequence to discover more accurate user interests, then design a unified model to solve the two tasks in a unified way. Next, we define the new problem to be handled.

Recall that we focus on the information content domain, let us formulate a user's behaviors with notations. On an information content service platform with both search engine and recommendation engine, the user $u$ could browse articles $B$ in the recommendation system, issue queries $Q$ to seek for information and click satisfied documents $D$ in the search engine. All these behaviors are sequential, so we integrate them into a heterogeneous behavior sequence in chronological order. Referring to existing session segmentation methods~\cite{Ge_2018_HRNN,Lu_2019_PSGAN}, we divide the user's whole behavior sequence into several \textbf{sessions} with 30 minutes of inactivity as the interval. Past behaviors in the current session are viewed as the \textbf{short-term} history. The other previous sessions constitute the \textbf{long-term} history. Specifically, we denote the user’s history sequence as $H=\{H^l, H^s\}=\{\{S_1,\ldots,S_{N-1}\}, S_N\}$, where $N$ is the number of sessions. Each session $S_i$ corresponds to a sub-sequence with both behaviors, such as $\{B_1,B_2,(Q_3,D^3_1,D^3_2),\ldots,\}$. 

We illustrate the whole behavior sequence in Figure~\ref{fig:behavior_sequence}. The horizontal edges indicate the sequential relationship between two consecutive actions, while the slanted edges point to the documents clicked under the corresponding query. The blue vertical lines separate sessions. For example, in the current Session $N$, the user first browses two articles in the recommendation system. Then, she enters a query in the search engine and clicks a document under this query. At the current moment $t$, the user would perform a target behavior, either search with an issued query $Q_t$ or browsing. For both tasks, we are supposed to infer the user's intent and return a personalized document list. Due to the same paradigm, we regard the recommendation task as personalized search with an empty query, and complete the two tasks in a unified personalized ranking style. Facing $Q_t$ or an empty query, the model is required to return a personalized document list based on the query and the user interests learned from the user's integrated behavior sequence.

\begin{figure*}
    \centering
    \includegraphics[width=0.95\linewidth,height=8.6cm]{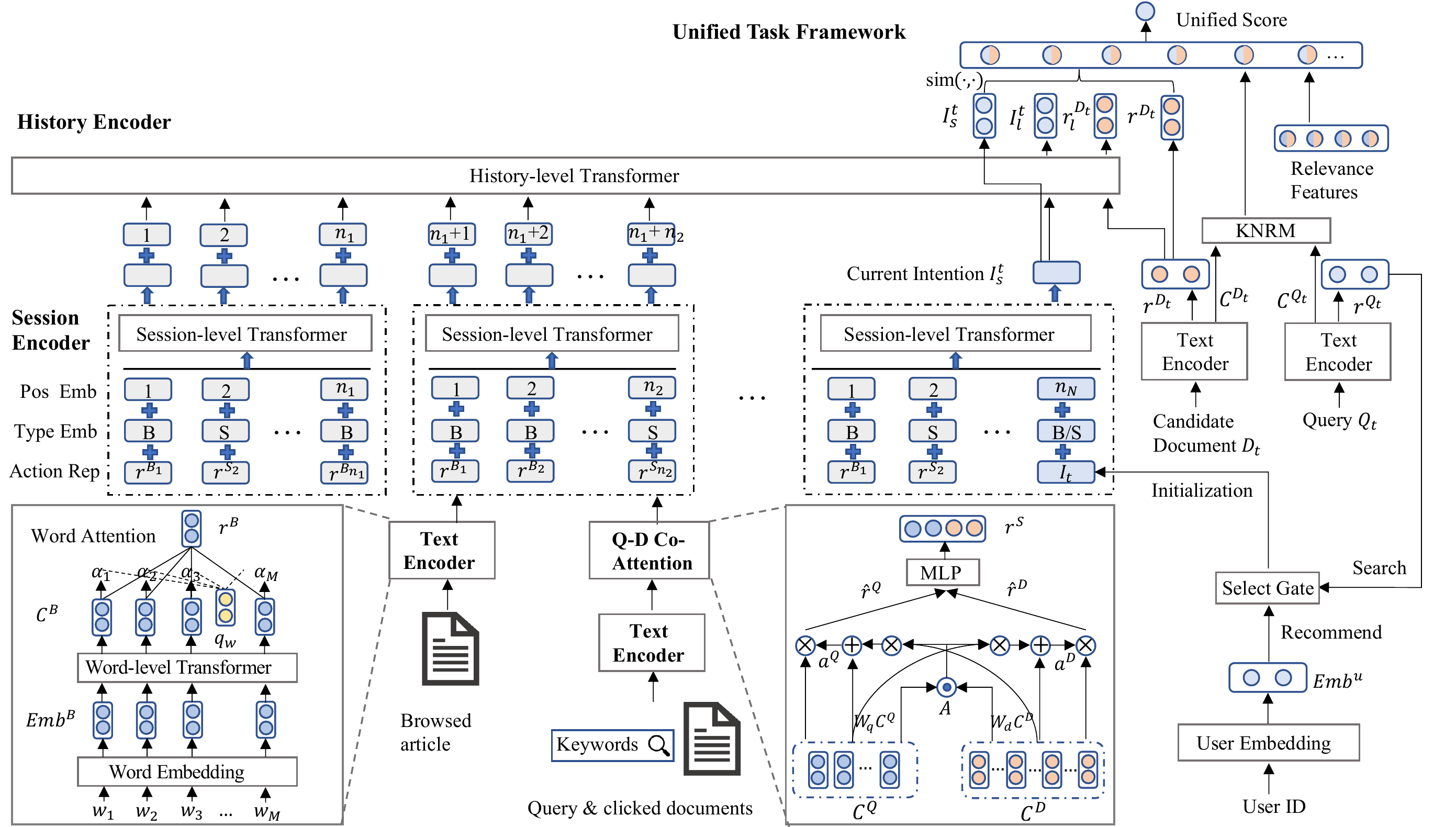}
    \caption{The architecture of our USER model. There are four major components: the text encoder to learn representations for queries and documents; the session encoder to model integrated behaviors in the current session; the history encoder to mine information from the long-term behavior sequence; the unified task framework to complete both tasks in a unified way. }
    \label{fig:model}
\end{figure*}

\section{USER: The Unified Model}\label{sec:model}
The architecture of our USER model is shown in Figure~\ref{fig:model}. First, the text encoder is used to learn representations for documents and queries. Second, the session encoder models the user's integrated behavior sequence within the current session to clarify her information need. Then, the history encoder enhances the user's intent representation by mining information from the long-term history. Finally, we design a unified task framework to complete personalized search and recommendation in a unified way. We present the details of each module in the remaining parts of this section.

\subsection{Text Encoder}\label{sec:text_encoder}
For each query $Q$, clicked document $D$ and browsed article $B$, we apply the text encoder to learn their semantic representations. Taking the calculation of a browsed article $B$ as an example, $B=[w_1,w_2,\ldots,w_M]$ where $M$ is the number of words in the article, the complete text encoder can be divided into three sub-layers. The first is the word embedding layer that converts the word sequence into a matrix with word vectors, i.e. $\text{Emb}^B=[v_1,v_2,\ldots,v_M] \in R^{dim\times M}$. $v_i$ corresponds to the low-dimensional word vector of $w_i$. In addition, contexts within the article are also helpful for users to figure out the true meaning of a word. For example, the different meanings of ``Apple'' in ``Apple fruit'' and ``Apple company'' can be distinguished based on the different contextual words ``fruit'' and ``company''. Therefore, we set a word-level transformer~\cite{Vaswani_2017_Tranformer} as the second sub-layer to obtain the context-aware word representations $C^B\in R^{dim\times M}$ by capturing interactions between words.
\begin{align}
    C^B &= \text{Transformer}_w(\text{Emb}^B).
\end{align}
The details about transformer can be referred to~\cite{Vaswani_2017_Tranformer}.

The last sub-layer is a word-level attention layer. In a piece of text, different words contribute different informativeness for expressing the semantics of this text. For instance, in the sequence `symptoms of novel coronavirus pneumonia', the word `symptoms' is very informative for learning the text representation, while `of' has little information. To highlight important words in a text sequence, we exploit a word-level attention mechanism to give them larger weights. We set a trainable vector $q_w$ as the query in the attention mechanism. The weights $\alpha\in R^M$ for all words are computed as:
\begin{align}
    \alpha = \text{softmax}(q_w^T\text{tanh}(W_v \times C^B + b_v)),
\end{align}
where $W_v$ and $b_v$ are parameters. The final contextual representation of the browsed document $r^B\in R^{dim\times 1}$ is the weighted sum of all the word vectors, i.e. $r^{B} = \sum_{i=1}^{M}\alpha_iC^B_i.$

Contextual representations of the query $r^Q$ and clicked document $r^D$ are computed in the same way.

\subsection{Session Encoder}
At the current time $t$, the user $u$ has a target action, either search or browsing. We represent her intention with a vector $I_t$. If the user issues a query $Q_t$ for search, the intention $I_t$ is initialized with the text representation of this query $r^{Q_t}$ computed by the text encoder. Otherwise, we use the corresponding trainable user embedding $\text{Emb}^u$ as initialization. This step is realized by a select gate, as:
\begin{align}
    I_t = 
    \begin{cases}
    r^{Q_t} & \text{if the target behavior is search} \\
    \text{Emb}^u & \text{if the target behavior is browsing}\\
    \end{cases}
\end{align}
Then, we mine information from the user's history comprised of search and browsing behaviors to clarify her personal intent $I_t$.

According to existing studies~\cite{Zhou_2020_HTPS,Ge_2018_HRNN}, it is thought that behaviors within a session show consistency in the user's information need. Thus, the user's past behaviors during the current session could provide rich contextual information for deducing her current intention. In the unified search \& recommendation scenario we study, there are both search and browsing actions in a session, as shown in Figure~\ref{fig:behavior_sequence}. We analyze that several possible relationships exist between the behaviors in the heterogeneous sequence: (1) For a document clicked under a query, we think this document satisfies the user's information need to be expressed by this query. It shows strong relevance between the query and the document. (2) After the user browses a series of recommended articles, she might be triggered to seek for more related information through proactive searches. (3) Queries are actively issued by the user, explicitly showing her preferences. With these queries and clicked documents, we can figure out the points of interest the user focuses on when browsing articles. We design a session encoder to capture these associations in the current session and employ the session context to enhance the intent representation.

First, for a historical query and the corresponding clicked documents, we are supposed to learn the strong relevance between them. Clicked documents indicate the user's intention contained in the query keywords, and the query highlights the important words in the documents. Thus, we suggest adopting the co-attention structure~\cite{Shu_2019_CoAttention} to calculate their representation vectors by fusing their interactive information, instead of the vanilla word-attention mechanism. Taking a query $Q$ and the clicked documents ${D_1,D_2,\ldots}$ as an example, the detailed computing process is as follows. At the first step, we obtain the contextual vector matrices $C^Q$ and $C^{D_i}$ for the query and each document through the word embedding layer and the word-level transformer of our text encoder. Vectors of all clicked documents are concatenated together as $C^D=[C^{D_1}; C^{D_2};\ldots]$. Then, we compute an affinity matrix $A$ between $C^Q$ and $C^D$.
\begin{align}
    A = \text{tanh}((C^Q)^TW_lC^D), 
\end{align}
where $W_l \in R^{dim\times dim}$ is a weight matrix to be learned. The attention weights for the query and documents are calculated based on the interactive features in the affinity matrix, as:
\begin{align}
    H^Q &= \tanh(W_qC^Q + (W_dC^D)A),\quad a^Q = \text{softmax}(W_{hq}^TH^Q),\\
    H^D &= \tanh(W_dC^D + (W_qC^Q)A^T),\quad a^D = \text{softmax}(W_{hd}^TH^D).
    % a^Q &= \text{softmax}(W_{hq}^TH^Q),\\
    % a^D &= \text{softmax}(W_{hd}^TH^D).
\end{align}
$W_q, W_d, W_{hq}, W_{hd}$ are parameters. $a^Q$ and $a^D$ are the attention weights for query keywords and document terms respectively. We calculate the attended representation for the query and documents as the weighted sum of the contextual vectors $C^Q$ and $C^D$.
\begin{align}
    \hat{r}^Q = \sum_{i=1}^{M} a^Q_iC^Q_i,\qquad %\\
    \hat{r}^D = \sum_{i=1}^{M} a^D_iC^D_i.
\end{align}
The two vectors are concatenated to generate the representation of a historical search behavior $r^S$ through an MLP layer, i.e. $r^S=\text{MLP}([r^Q;r^D])$. For a browsing behavior made in recommendation, it corresponds to only a browsed article $B$. Thus, its representation is just the article representation $r^B$ calculated by the text encoder. %\dou{same lengths?}

With the representation of all past behaviors in the current session calculated, $H^s=\{r^{B_1}, r^{S_2}, \ldots\}$, we could capture the relationships between the search and browsing behaviors, and fuse the session context into the user's current intention. We combine $H^s$ with the target intention $I_t$ and pass them through a session-level transformer for interaction. On account of the behaviors are sequential and heterogeneous, we add the position and type information of each behavior for clarification. The action type includes search (S) and browsing (B). Finally, the output of the last position $I_t^s$ represents the user's current intention fusing the session context.
\begin{align}
    I_t^s = \text{Transformer}_s^{\text{last}}([H^s,I_t] + [H^s,I_t]_P + [H^s,I_t]_T).
\end{align}
$[H^s,I_t]_P$, $[H^s,I_t]_T$ are the position embedding and type embedding. $\text{Transformer}_s^{\text{last}}(\cdot)$ means taking the output of the last position.

\subsection{History Encoder}
With the session encoder described above, we clarify the user's current information need under the help of the short-term history, obtaining $I_t^s$. But for the situation with little session history, it is still ambiguous due to the lack of session context. The user's long-term behavior history often reflects relatively stable interests, which also provides some assistant information. Thus, we further model the long-term history to enhance the user's intent representation based on $I_t^s$. At first, we process each historical session with the session encoder to capture the connections between search and browsing behaviors, getting the contextual representation for all historical behaviors, $H^l=[\{r_s^{B_1}, r_s^{S_2}, \ldots\},\{r_s^{B_1}, \ldots\},\ldots]$. We concatenate all session sub-sequences as a long behavior sequence and combine it with the target action as $[H^l, I_t^s]$. Then, a history-level transformer module is conducted on the long-term heterogeneous sequence to fuse the history information into the current intention. To preserve the sequential information between actions, we involve the position of each behavior $\{1,2,\ldots\,t\}$. In the final, we take the output of the last position as the user's intent representation enhanced by the long-term behavior history, denoted as $I_t^l$. %\dou{ambiguous: idx of the behavior, or session?}\jing{behavior}
\begin{align}
    I_t^l = \text{Transformer}_h^{\text{last}}([H^l,I_t^s] + [H^l,I_t^s]_P),
\end{align}
where $[H^l,I_t^s]_P$ is the position embedding.

Motivated by some news recommendation models~\cite{Wu_2019_NPA,Wu_2019_NRMS}, the user's attention to a document is also impacted by her interests. Besides, the user might intend to find a specific document that appeared in the history, as analyzed in~\cite{Zhou_2020_RPMN}. Thus, for the candidate document $D_t$, we can use the long-term history to enhance its representation $r^{D_t}$ calculated by the text encoder in the same way as the target intent, getting $r^{D_t}_l$
\begin{align}
    r^{D_t}_l = \text{Transformer}_h^{\text{last}}([H^l,r^{D_t}] + [H^l,r^{D_t}]_P).
\end{align}
We will use $r^{D_t}_l$ together with $r^{D_t}$ to calculate the personalized ranking score for the candidate document in the unified task framework that will be introduced in the next part.

\subsection{Unified Task Framework}
As for the personalized search and recommendation tasks in the information content domain, the main difference between them is whether there is an issued query. In the problem definition, we claim to unify the two different tasks as a unified problem by regarding the recommendation task as personalized search with an empty query. We represent the user's current intention as $I_t$ that is initialized with the issued query $Q_t$ for search or the user embedding $\text{Emb}^u$ for recommendation. The unified problem is to rank the candidate document $D_t$ based on the personalized relevance that is calculated with the current intention $I_t$, the query $Q_t$ (empty for recommendation) and the user history $H$. The personalized relevance is denoted as $p^{\texttt{unified}}(D_t|I_t,Q_t,H)$.

Through the text encoder, session encoder and history encoder, we get the representations of the user's current intention and candidate document, i.e. $I_t^s$, $I_t^l$, $r^{D_t}$ and $r^{D_t}_l$. We calculate the relevance between each pair of them by cosine similarity $\text{sim}(\cdot,\cdot)$. Moreover, for the personalized search task, the correlation between the candidate document and the query keywords is also critical. Thus, we additionally pay attention to the interactive features between the context-aware representations of the query and document, i.e. $C^{Q_t}$ and $C^{D_t}$. We exploit the interaction-based component KNRM~\cite{Xiong_2017_KNRM} to calculate the interactive score $\text{inter}(C^{Q_t}, C^{D_t})$. The detailed calculation process can be found in~\cite{Xiong_2017_KNRM}. Besides, following~\cite{Ge_2018_HRNN,Lu_2019_PSGAN}, we also extract several relevance-based features $F_{q,d}$ for personalized search. When calculating the relevance for articles in recommendation, the interaction score and features are all empty. Finally, the score for the candidate document is calculated by aggregating all these scores and features with an MLP layer, as: 
\begin{align}
    f^{\texttt{unified}} & = [\text{sim}(I_t^s,r^{D_t}),\text{sim}(I_t^l,r^{D_t}),\text{sim}(I_t^s,r^{D_t}_l),\nonumber \\
    & \text{sim}(I_t^l,r^{D_t}_l), \text{inter}(C^{Q_t}, C^{D_t}),F_{q,d}],\\
    p^{\texttt{unified}}&(D_t|I_t,Q_t,H) = \Phi(f^{\texttt{unified}}).
\end{align}
$\Phi()$ represents an MLP layer without an activation function. Whether for the search or recommendation task, we generate personalized document list by calculating relevance scores in this way.

\begin{table}[!t]
  \caption{Basic statistics of the dataset.}
  \label{tab:statistics}
  \begin{tabular}{p{0.08\textwidth}p{0.06\textwidth}||p{0.15\textwidth}p{0.06\textwidth}}
   \toprule
    Item & Statistic & Item & Statistic \\
   \midrule
   \#days & 92 & \#search records & 203,190 \\
    \#users & 100,000 & \#recommend records & 840,867 \\
   \#sessions & 515,247 & avg. session length & 3.58 \\ 
   \bottomrule
  \end{tabular}
\end{table}

\subsection{Training and Optimization}
We adopt a pairwise manner to train our USER model. For both personalized search and recommendation tasks, we construct each training sample as a document group comprised of a positive document and $K$ negative documents presented in the same impression, represented as $\{D^+,(D_1^-,\ldots,D_K^-)\}$. For each document group, we aim to maximize the score of the positive document and minimize that of those negative documents. The loss $\mathcal{L}$ is computed as the negative log-likelihood of the positive sample. We have:
\begin{align}
    \mathcal{L} &= -\log( \frac{\exp(p^{\texttt{unified}}(D^+))}{\exp(p^{\texttt{unified}}(D^+)) + \sum_{i=1}^K\exp(p^{\texttt{unified}}(D_i^-))}).
\end{align}
where $p^{\texttt{unified}}(\cdot)$ is the abbreviation of $p^{\texttt{unified}}(\cdot|I_t,Q_t,H)$. We minimize the loss with the Adam optimizer. 

In the unified scenario, we have access to both search and recommendation data. Thus, we can train one USER model with data from the two tasks and apply the trained model to both of them. However, there may be a problem that some gaps exist between the data distributions of the search task and recommendation task. The only unified model trained on the data from the two tasks is difficult to achieve the best performance on both of them. Therefore, we propose an alternative training method. We first pre-train a unified model with both task data. Then, we make a copy for each task and finetune it with the corresponding task data to fit the individual data distribution. In this case, the model not only benefits from more training data but also adapts to the specific task.

\section{Experimental Settings}\label{sec:settings}
\subsection{Dataset and Metrics}
\textbf{Dataset}
There is no public dataset with both search and recommendation logs of a shared set of users in the information content domain. To evaluate our unified model, we construct a dataset comprised of users' search and browsing behaviors from a popular information service platform that has both search and recommendation engines. We randomly sample 100,000 users. Then, we obtain their search logs in its search engine and browsed articles recommended by the recommendation system for three months. The whole log is preprocessed via data masking to protect user privacy.

Each piece of search data includes an anonymous user ID, the action time, a query, top 20 returned documents, click tags and click dwelling time. As for each recommendation record, only a browsed article is kept, without other presented but unclicked documents. We generate pseudo unclicked documents for each browsed article for model training. We rank all documents in the recommendation log based on a weighted score of the popularity measured by the click count and the topic similarity with the browsed article calculated by cosine similarity. Nine negative documents ranked at the top are sampled for each browsed article. The original recommendation list is randomly shuffled. All search and browsing behaviors of each user are merged into a sequence in chronological order.

We separate a user’s whole behavior sequence into sessions with 30 minutes of inactivity as the interval~\cite{Ge_2018_HRNN,Lu_2019_PSGAN}. Users' browsing behaviors are usually more frequent than search behaviors, which leads to an unbalance in the dataset. Since we intend to explore the relatedness between the search and browsing behaviors, we sample sessions containing both actions and three sessions before and after these sessions. To guarantee each user has enough history for building user profile, we treat the log data of the first eight weeks as the historical set and the other five weeks log as the experimental data. The experimental data is used for training, validation and testing with 4:1:1 ratio. The statistics are shown in Table~\ref{tab:statistics}.

\textbf{Metrics}
Referring to existing works~\cite{Wu_2019_NPA,Wu_2019_NRMS}, the recommendation task is also to re-rank the candidate documents. For both tasks, we take the sat-clicked documents with more than 30 seconds of dwelling time as relevant and the others as irrelevant.
We choose common ranking metrics to evaluate our model and baselines, including MAP, MRR, P@1, Avg.C (average position of the clicked documents), NDCG@5 and NDCG@10. For recommendation, we also adopt AUC to measure the click-through rate.

\begin{table*}[t]
 \center
 \setlength{\abovecaptionskip}{0.1cm}
 \setlength{\belowcaptionskip}{0.1cm}
 \caption{Overall performance in various scenarios: personalized search with only search data, recommend with browsing data, unified search \& recommend with both data. The relative percentages are computed based on the original rankings. The best results in each scenario are shown with underlines. The overall best results are in bold. "$\dagger$" indicates significant improvements over all corresponding baselines, with paired t-test at p $<$ 0.05 level. MRR is used for search and AUC for recommendation.}
  \label{tab:overall performance}
  \begin{tabular}{p{0.10\textwidth}|p{0.08\textwidth}|p{0.04\textwidth}p{0.047\textwidth}|p{0.04\textwidth}p{0.047\textwidth}|p{0.04\textwidth}p{0.048\textwidth}|p{0.04\textwidth}p{0.047\textwidth}|p{0.04\textwidth}p{0.047\textwidth}|p{0.04\textwidth}p{0.047\textwidth}}
       \toprule
       Scenarios & Models & \multicolumn{2}{c|}{MAP} & \multicolumn{2}{c|}{MRR/AUC} & \multicolumn{2}{c|}{P@1} & \multicolumn{2}{c|}{Avg.C} & \multicolumn{2}{c|}{NDCG@5} & \multicolumn{2}{c}{NDCG@10}  \\ \hline
       \multirow{6}*{\shortstack{Personalized \\ Search}} & Ori. & .6614 & - & .6859 & - & .5766 & - & 3.898 & - & .6720 & - & .7148 & - \\
       & HRNN & .6707 & +1.41\% & .6951 & +1.34\% & .5800 & +0.59\% & 3.727 & +4.39\% & .6839 & +1.77\% & .7258 & +1.54\% \\ 
       & RPMN & .6724 & +1.66\% & .6962 & +1.50\% & .5795 & +0.50\% & 3.642 & +6.57\% & .6859 & +2.07\% & .7296 & +2.07\% \\ 
       & PEPS & .6727 & +1.71\% & .6979 & +1.75\% & .5811 & +0.78\% & 3.626 & +6.98\% & .6867 & +2.19\% & .7303 & +2.17\% \\ 
       & HTPS & .6749 & +2.04\% & .6991 & +1.92\% & .5821 & +0.95\% & 3.615 & +7.26\% & .6894 & +2.59\% & .7326 & +2.49\% \\ 
       & USER-S & \underline{.6768} & +2.33\% & \underline{.7008} & +2.17\% & \underline{.5831} & +1.13\% & \underline{3.564} & +8.57\% & \underline{.6909} & +2.81\% & \underline{.7341} & +2.70\% \\ \hline
       \multirow{6}*{\shortstack{Recommend}} & Ori. & .2928 & - & .4994 & - & .0989 & - & 5.493 & - & .2952 & - & .4543 & - \\
       & NPA & .4787 & +63.49\% & .7104 & +42.25\% & .2679 & +170.9\% & 3.489 & +36.48\% & .5306 & +79.74\% & .6037 & +32.89\% \\
       & LSTUR & .4841 & +65.33\% & .7131 & +42.79\% & .2767 & +179.8\% & 3.564 & +35.12\% & .5268 & +78.46\% & .6075 & +33.72\% \\ 
       & NRMS & .4902 & +67.42\% & .7124 & +42.65\% & .2794 & +182.5\% & 3.584 & +34.75\% & .5263 & +78.29\% & .6128 & +34.89\% \\ 
       & GERL & .4591 & +56.80\% & .7229 & +44.75\% & .2417 & +144.4\% & 3.499 & +36.30\% & .5253 & +77.95\% & .5885 & +29.54\% \\ 
       & USER-R & \underline{.4975}$^\dagger$ & +69.91\% & \underline{.7326}$^\dagger$ & +46.70\% & \underline{.2825}$^\dagger$ & +185.6\% & \underline{3.381}$^\dagger$ & +38.45\% & \underline{.5386}$^\dagger$ & +82.45\% & \underline{.6204}$^\dagger$ & +36.56\% \\ \hline
       \multirow{8}*{\shortstack{Unified Search\\ \& Recommend}} & \multicolumn{7}{l}{Personalized Search Task} \\ \cline{2-14}
       % & HRNN & 0.00 & 0.00 & 0.00 & 0.00 & 0.00 & 0.00 & 0.00 & 0.00 & 0.00 & 0.00\\ 
       & HTPS & .6789 & +2.65\% & .7012 & +2.23\% & .5851 & +1.47\% & 3.546 & +9.03\% & .6941 & +3.29\% & .7369 & +3.09\% \\ 
       & JSR-HRNN & .6705 & +1.38\% & .6954 & +1.39\% & .5796 & +0.52\% & 3.724 & +4.46\% & .6843 & +1.83\% & .7258 & +1.54\% \\ 
       & USER & \textbf{\underline{.6845}}$^\dagger$ & +3.49\% & \textbf{\underline{.7074}}$^\dagger$ & +3.13\% & \textbf{\underline{.5892}}$^\dagger$ & +2.19\% & \textbf{\underline{3.479}}$^\dagger$ & +10.75\% & \textbf{\underline{.7001}}$^\dagger$ & +4.18\% & \textbf{\underline{.7422}}$^\dagger$ & +3.83\% \\ \cline{2-14}
       & \multicolumn{7}{l}{Recommendation Task} \\ \cline{2-14}
       & NRMS & .4984 & +70.22\% & .7203 & +44.23\% & .2948 & +187.9\% & 3.496 & +36.35\% & .5361 & +81.61\% & .6147 & +35.31\% \\ 
       & JSR-NRMS & .4931 & +68.41\% & .7251 & +45.19\% & .2765 & +179.6\% & 3.386 & +38.36\% & .5432 & +84.01\% & .6143 & +35.22\% \\ 
      & USER & \textbf{\underline{.5035}}$^\dagger$ & +71.96\% & \textbf{\underline{.7442}}$^\dagger$ & +49.02\% & \textbf{\underline{.2910}}$^\dagger$ & +194.2\% & \textbf{\underline{3.258}}$^\dagger$ & +40.69\% & \textbf{\underline{.5514}}$^\dagger$ & +86.79\% & \textbf{\underline{.6222}}$^\dagger$ & +36.96\%\\
       \bottomrule
  \end{tabular}
\end{table*}

\subsection{Baselines}
The original search results are returned by the search engine. The original recommendation lists are randomly shuffled. Besides, we compare our model with state-of-the-art personalized search models, news recommendation models and the joint framework~\cite{Zamani_2018_JSR}. % Details are listed as follows.

\textbf{HRNN}~\cite{Ge_2018_HRNN}: A hierarchical RNN model with query-aware attention to dynamically mine relevant history information.

\textbf{RPMN}~\cite{Zhou_2020_RPMN}: This model captures complex re-finding patterns of previous queries or documents with the memory network.

\textbf{PEPS}~\cite{Yao_2020_PEPS}: Yao et al. claim that different users have different understandings of the same word due to their knowledge. They learn personal word embeddings to clarify the query keywords.

\textbf{HTPS}~\cite{Zhou_2020_HTPS}: It encodes the user's history as the context information to disambiguate the current query. We adapt it to the unified scenario by adding the user's browsed articles into her history.

\textbf{NPA}~\cite{Wu_2019_NPA}: The model sets user embeddings to compute personalized word- and news-level attention. It highlights important words and articles to generate informative news and user representations.

\textbf{NRMS}~\cite{Wu_2019_NRMS}: This model utilizes multi-head self-attention to learn news and user representations by capturing the relatedness between words and browsed articles. By adding documents clicked in the search history, we adapt it into the unified scenario.

\textbf{LSTUR}~\cite{An_2019_LSTUR}: It includes the short-term user interests modeled from the recent clicked articles with GRU and the long-term profile corresponding to a trainable user embedding.

\textbf{GERL}~\cite{Ge_2020_GERL}: It applies transformer on the user's interaction graph to capture high-order associations between users and news.

\textbf{JSR}~\cite{Zamani_2018_JSR}: This is a general joint framework that trains a separate search model and a recommendation model by optimizing a joint loss. We select HRNN for search and NRMS for recommendation.

\textbf{USER}: This is the unified model proposed in this paper. \textbf{USER-S} and \textbf{USER-R} are the variants used in independent search and recommendation scenarios respectively. They share the same structure as \textbf{USER} but have access to only the data of that single task.

% \subsection{Model Settings}
We conduct multiple sets of experiments to decide the model parameters as follows\footnote{The code is on https://github.com/jingjyyao/Personalized-Search/tree/main/USER}. The size of word embeddings, pretrained by word2vec on all logs, and user embeddings is 100. Due to users' click decisions are usually made based on titles, we use titles in our experiment, instead of complete articles. For each query or document title, the max sequence length is 30. In the history sequence, we maintain up to 20 sessions and the maximum number of user behaviors in a session is 5. The number of heads in the transformer is 8 and the hidden dimension is 50. The number $K$ of negative samples in each document group is 4. The learning rate is $1e-3$.

\begin{table*}[t]
 \center
 \setlength{\abovecaptionskip}{0.1cm}
 \setlength{\belowcaptionskip}{0.1cm}
 \caption{Performance of USER variants. The relative percentages are calculated based on the complete USER model.}
  \label{tab:ablation study}
  \begin{tabular}{p{0.16\textwidth}|p{0.04\textwidth}p{0.048\textwidth}|p{0.04\textwidth}p{0.048\textwidth}|p{0.04\textwidth}p{0.048\textwidth}|p{0.04\textwidth}p{0.055\textwidth}|p{0.04\textwidth}p{0.055\textwidth}|p{0.04\textwidth}p{0.055\textwidth}}
       \toprule
       \multirow{2}*{Variants} & \multicolumn{6}{c|}{Personalized Search} & \multicolumn{6}{c}{Recommendation} \\ \cline{2-13}
       & \multicolumn{2}{c|}{MAP} & \multicolumn{2}{c|}{NDCG@5} & \multicolumn{2}{c|}{NDCG@10} & \multicolumn{2}{c|}{MAP} & \multicolumn{2}{c|}{NDCG@5} & \multicolumn{2}{c}{NDCG@10}  \\ \hline
       % \multicolumn{7}{l}{Sequence Encoders} \\ \cline{1-13}
       USER & \textbf{.6845} & - & \textbf{.7001} & - & \textbf{.7422} & - & \textbf{.5035} & - & \textbf{.5514} & - & \textbf{.6222} & -\\ \hline
       \;\;w/o Session Encoder & .6760 & -1.24\% & .6934 & -0.96\% & .7343 & -1.06\% & .3963 & -21.29\% & .4504 & -18.32\% & .5393 & -13.32\%\\ 
       \;\;w/o History Encoder & .6768 & -1.12\% & .6937 & -0.91\% & .7348 & -1.00\% & .4443 & -11.76\% & .5026 & -8.85\% & .5766 & -7.33\%\\ \hline
       % \multicolumn{7}{l}{Unified Task} \\ \cline{1-13}
       % w/o Search Data & - & - & - & - & - & - & .4975 & -1.19\% & .5386 & -2.32\% & .6204 & -0.29\%\\  
       % w/o Recommend Data & .6768 & -0.92\% & .6909 & -1.31\% & .7341 & -0.89\% & - & - & - & - & - & - \\
       \;\;w/o Unified Pre-train & .6801 & -0.64\% & .6953 & -0.69\% & .7383 & -0.53\% & .5032 & -0.06\% & .5510 & -0.07\% & .6215 & -0.11\% \\
       \;\;w/o Unified Data & .6768 & -1.12\% & .6909 & -1.31\% & .7341 & -1.09\% & .4975 & -1.19\% & .5386 & -2.32\% & .6204 & -0.29\%\\
       \bottomrule
  \end{tabular}
\end{table*}
\section{Experimental Results}\label{sec:results}
\subsection{Overall Results}
We compare all models in various scenarios: pure personalized search with only search data, pure recommendation with only recommendation data and unified scenario with both data. The results are shown in Table~\ref{tab:overall performance}. We have several findings:

(1) The comparison of the same model trained with the independent task dataset and the unified dataset. \textbf{For HTPS, NRMS and USER, their performance on the unified dataset is better than that on the independent task data.} For example, the personalized search model HTPS trained on the unified history promotes 0.6\% in MAP based on that trained with pure search data. The recommendation model NRMS has 1.6\% improvement in MAP with the unified data. Consistently, our USER model in the unified scenario also shows improvements over USER-S and USER-R on all metrics. Compared to the independent task data, the unified dataset is comprised of both search and browsing behaviors, from which we analyze the user's preferences. The results demonstrate that a more precise user interest profile can be constructed based on the integrated behavior sequence to improve ranking qualities.

(2) The comparison of our USER model and the separate personalized search or recommendation baselines. \textbf{The USER-S and USER-R variants achieve better results than the corresponding baselines on independent scenarios. Greater improvements are observed on the complete USER model in the unified case with both data, with paired t-test at p<0.05 level.} Specifically, on the pure personalized search, USER-S outperforms the HTPS. In recommendation, USER-R promotes NRMS on all evaluation metrics. This proves that our history encoders can effectively learn user interests to improve personalized rankings. Furthermore, the complete USER model promotes HTPS more greatly in the unified scenario. We analyze it may because the USER model is pre-trained by both search and recommendation tasks on the unified dataset, which benefits from more training data.

(3) The comparison of our unified model USER and the general joint framework JSR. \textbf{Compared to JSR, USER improves the corresponding separate variants (USER-S and USER-R) much better by training with the unified data.} The HRNN and NRMS combined in JSR show similar performance to the original HRNN and NRMS. However, the USER model achieves 1.14\% improvements in MAP over the USER-S and 1.20\% in MAP over the USER-R. JSR simply combines a personalized search model and a recommendation model through optimizing a joint loss, without exploiting any interactions between them. In the USER model, we integrate the two kinds of behaviors into a heterogeneous sequence and complete both tasks based on this sequence. The results suggest that USER provides a better approach to aggregate the two tasks and capture the associations between them to promote each other.

To conclude, \textbf{with the unified data comprised of the user's search and browsing logs, a more comprehensive user profile and more training samples can be obtained for personalization. Besides, the USER model is promising to capture the relatedness between the two tasks to promote each other.}
\subsection{Ablation Study}
To analyze how the major modules in our model impact the effects, we conduct several ablation studies. The variants are as follows.
%
%We conduct ablation studies to analyze major modules.

\textbf{USER w/o Session Encoder}: We discard the short-term history and the session encoder for clarification.

\textbf{USER w/o History Encoder}: In this variant, we remove the long-term history and the history-level transformer.

\textbf{USER w/o Unified Pre-train}: We skip pre-training one unified model with the training data from both tasks, but train two separate models from scratch, with integrated history sequences.

\textbf{USER w/o Unified Data}: With only separate task data not the unified dataset, USER degrades to USER-S and USER-R respectively.

From the results shown in Table~\ref{tab:ablation study}, we can observe that:

(1) Removing the session encoder or history encoder and the corresponding behavior history causes a decline in all evaluation metrics for both personalized search and recommendation tasks. This proves that both encoders mine information from the user's history to help personalization. The session encoder captures the user’s consistent intention in the current session. The history encoder learns stable user interests in the long-term history. The two parts help clarify the user's current information need together.

(2) There is a decrease in the ranking results when skipping the unified pre-training, especially for the personalized search task. This confirms the benefits of more training samples constructed from both task data in our unified model. It has few impacts on the recommendation task. A possible reason is that the browsing behaviors in recommendation are usually far more frequent than search behaviors, thus the recommendation task has enough training samples. Discarding the unified data leads to a greater decline in both tasks. On a separate task dataset, only one kind of user behavior is available whether in history or training. This decline demonstrates that the integrated behavior sequence is more informative.

\begin{figure}
    \centering
    \includegraphics[width=0.98\linewidth]{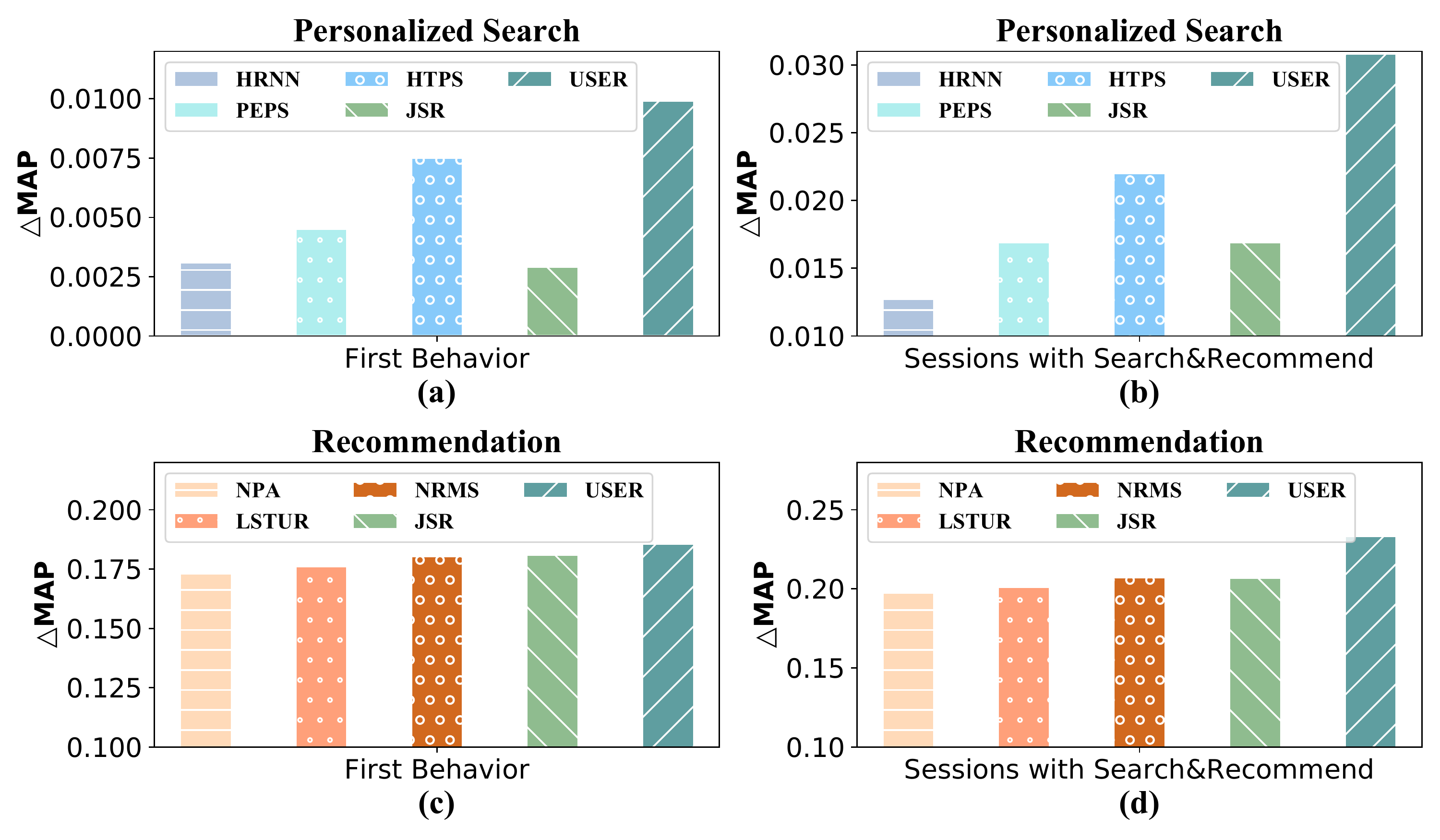}
    \caption{Performance on different data subsets. (a), (c): the first search/recommend behavior of each user; (b), (d): sessions with both search \& recommendation.}
    \label{fig:different_set}
\end{figure}

\begin{figure*}
    \centering
    \includegraphics[width=0.80\linewidth,height=4.3cm]{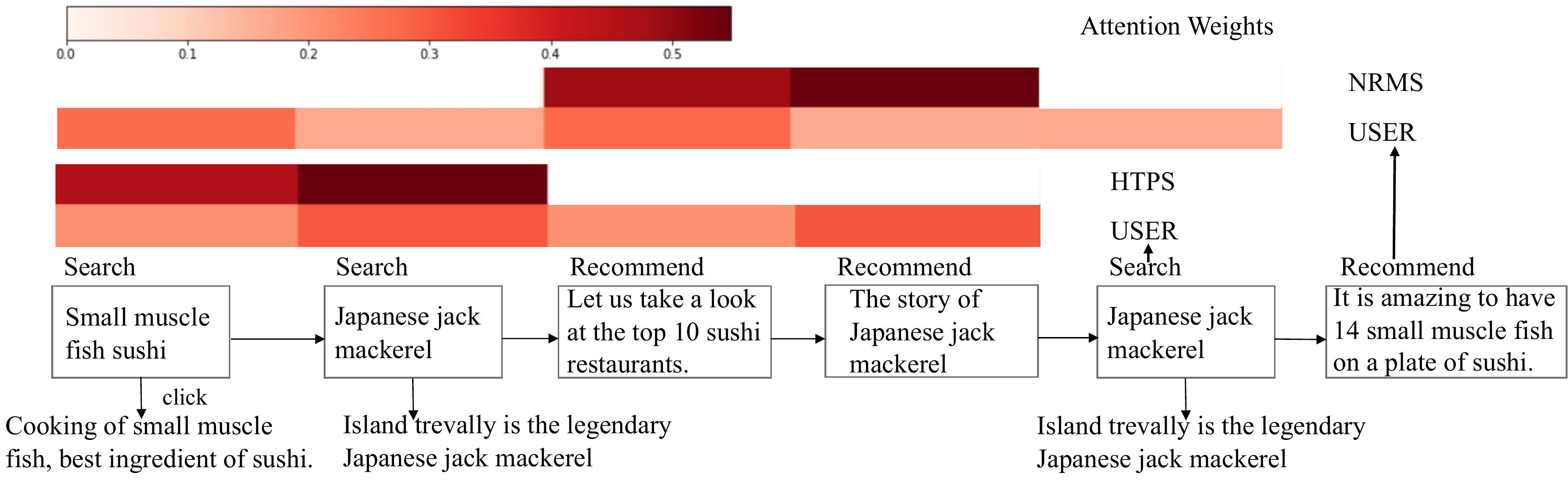}
    \caption{Illustration of the heterogeneous behavior sequence in a session and the attention weights of the current action over historical behaviors in different models. A darker area indicates a larger weight.}
    \label{fig:case_study}
\end{figure*}

\subsection{Performance on Specific Set}
We further test our model and baselines on different subsets: the first search/recommend behavior of each user, and sessions with search \& recommendation. The results are shown in Figure~\ref{fig:different_set}, using the improvement of MAP over the original ranking as the metric.

\textbf{First Search/Recommend Behavior.} We claim that USER is promising to alleviate data sparsity by merging the user's search and recommendation logs. To verify this effectiveness of USER, we sample each user's first search record and recommendation record in the testing data to construct a subset. In this subset, there is little search history for each piece of search data, and little browsing history for each recommendation record. It is a cold-start case for separate personalized search and recommendation tasks. 
From Figure~\ref{fig:different_set} (a) and (c), we find that USER trained on the unified dataset outperforms the corresponding baselines with only separate search or recommendation data. In the unified situation, for the user's first search behavior with little search history, the browsing history can be a supplement for mining the user's preferences. As for the first recommendation sample, the search history can also be used as auxiliary information. Thus, we think that combining the two tasks as well as the corresponding behaviors indeed eliminates the problem of user data sparsity and the cold-start challenge.% faced by each single task.

\textbf{Sessions with Search \& Recommendation.} In this paper, we intend to explore the relatedness between the user’s search and browsing behaviors to promote the two tasks. Therefore, we sample a subset comprised of sessions with both behaviors. We select several independent baselines and JSR for comparison.

From Figure~\ref{fig:different_set} (b) and (d), we find that USER achieves the best on both tasks. The other joint model JSR that consists of HRNN and NRMS shows similar performance to the separate models. In USER model, we deduce the user's intent based on the integrated behavior sequence. Thus, the potential relatedness between the two kinds of behaviors can be captured to promote personalization, especially for these sessions with both behaviors. However, JSR trains two separate models through a joint loss, which might have difficulty learning the interactions between the two tasks. These results also suggest that USER copes with the unified scenario better than JSR.

\subsection{Case Study}
In this paper, we focus on the situation with both search and recommendation services in the information content domain. We design a unified model (USER) to jointly handle the two tasks. To illustrate the advantages of our model more intuitively, we conduct a case study to analyze the user's mixed behaviors within a session. Moreover, we discuss the impacts of the user's historical behaviors on the current action in USER, HTPS and NRMS. The impacts are indicated by the attention weights. The results are in Figure~\ref{fig:case_study}.

Observing the user's behaviors in the session, we find the user's preferences reflected by the search behaviors and browsing behaviors are consistent, probably about sushi, small muscle fish and Japanese jack mackerel. Besides, there is some relatedness between the two kinds of behaviors. For example, the user browses the article titled ``The story of Japanese jack mackerel'' in recommendation, followed by a query ``Japanese jack mackerel'' to 
seek more relevant information. Thus, integrating the two tasks together has the potential to promote each other. With the aggregated behaviors, we can mine more precise information about the user's interests to help the current ranking. Let us take the last search query ``Japanese jack mackerel’’ as an example. Obviously, this query is strongly relevant to both the historical query ``Japanese jack mackerel’’ and the browsed article ``The story of Japanese jack mackerel’’. USER pays high attention to both the two strongly relevant behaviors. However, HTPS, which is proposed for the independent search case, can only attend to the historical queries, without any information about the browsing actions. With regard to the last recommendation, the historical query ``small muscle fish’’ also reflects relevant user interests, which will be highlighted in USER. \textbf{This case study fully proves the value of aggregating the two separate tasks together and our proposal of the unified model}.
\section{Conclusion}\label{sec:conclusion}
In this paper, we focus on the connections between the personalized search and recommendation in the information content domain, and explore an effective approach to jointly model them together. We integrate the user's search and browsing behaviors into a heterogeneous behavior sequence. Then, we propose the unified model USER. It includes encoders to mine information from the heterogeneous behavior sequence for personalization and a unified task framework to solve both tasks in a unified ranking style. We experiment with a dataset comprised of both behaviors constructed from a real-world commercial platform. The results confirm that our model outperforms the state-of-the-art separate baselines on both tasks. In the future, we will combine the two tasks better.

\section*{Acknowledgements}
Zhicheng Dou is the corresponding author. This work was supported by National Natural Science Foundation of China No. 61872370 and No. 61832017,  Beijing Outstanding Young Scientist Program NO. BJJWZYJH012019100020098, Shandong Provincial Natural Science Foundation under Grant ZR2019ZD06, and Intelligent Social Governance Platform, Major Innovation \& Planning Interdisciplinary Platform for the ``Double-First Class'' Initiative, Renmin University of China. I also wish to acknowledge the support provided and contribution made by Public Policy and Decision-making Research Lab of Renmin University of China.

\newpage
%\bibliographystyle{ACM-Reference-Format}
%\bibliography{reference}
%%% -*-BibTeX-*-
%%% Do NOT edit. File created by BibTeX with style
%%% ACM-Reference-Format-Journals [18-Jan-2012].

\end{document}